\documentstyle[11pt,aaspp4,nick]{article}
\begin{document}


\title{Effect of Reionization on Structure Formation in the Universe}
\author{Nickolay Y.\ Gnedin}
\affil{CASA, University of Colorado, Boulder, CO 80309;
e-mail: \it gnedin@casa.colorado.edu}


\load{\scriptsize}{\sc}

\def\A{{\cal A}}
\def\B{{\cal B}}
\def\ion#1#2{\rm #1\,\sc #2}
\def\HI{{\ion{H}{i}}}
\def\HII{{\ion{H}{ii}}}
\def\GI{{\ion{He}{i}}}
\def\GII{{\ion{He}{ii}}}
\def\GIII{{\ion{He}{iii}}}
\def\MH{{{\rm H}_2}}
\def\Hp{{{\rm H}_2^+}}
\def\Hm{{{\rm H}^-}}

\def\dim#1{\mbox{\,#1}}

\def\figdir{.}
\def\placefig#1{#1}

\begin{abstract}
I use simulations of cosmological reionization to quantify
the effect of photoionization on the gas fraction in low mass
objects, in particular the characteristic mass scale below which
the gas fraction is reduced compared to the universal value.
I show that this characteristic scale can be up to an order of
magnitude lower than the linear theory Jeans mass, and that even
if one defines the Jeans mass at a higher overdensity,
it does not track the evolution of this characteristic suppression
mass.  Instead, the filtering mass, which corresponds directly 
to the scale over which baryonic perturbations are smoothed in linear
perturbation theory, provides a remarkably good fit to the characteristic 
mass scale.  Thus, it appears that the effect of reionization on structure
formation in both the linear and nonlinear regimes is described
by a single characteristic scale, the filtering scale of baryonic 
perturbations.  In contrast to the Jeans mass, the filtering mass depends
on the full thermal history of the gas instead of the instantaneous
value of the sound speed, so it accounts for the finite time required
for pressure to influence the gas distribution in the expanding universe.
In addition to the characteristic suppression mass, I study the full shape
of the probability distribution to find an object with a given gas mass 
among all the objects with the same total mass, and I show that the
numerical results can be described by a simple fitting formula that again
depends only on the filtering mass.  This simple description of the 
probability distribution may be useful for semi-analytical modeling of 
structure formation in the early universe.
\end{abstract}


\section{Introduction}

The effect of cosmological reionization on the formation and evolution of
low mass objects has been under the scrutiny of theorists for a long time,
ever since Ikeuchi (1986) and Rees (1986) independently pointed out that
the increase in the temperature of the cosmic gas during reionization will
suppress the formation of small galaxies with masses below the Jeans
mass. Several attempts to quantify the effect of reionization on 
low mass galaxies have been made since using semi-analytical calculations
(Babul \& Rees 1992; Efstathiou 1992; Shapiro, Giroux, \& Babul 1994), 
spherically symmetric modeling
(Haiman, Thoul, \& Loeb 1996; Thoul \& Weinberg 1996), and 
three-dimensional cosmological hydrodynamic simulations (Quinn, Katz, \&
Efstathiou 1996; Weinberg, Hernquist, \& Katz 1997; Navarro \& Steinmetz
1997). 

While confirming the general proposition that reionization
suppresses formation of low mass galaxies, these studies 
do not give the full description of the impact of reionization on
the gas fraction in the low mass objects. Particularly, one can expect
that the effect of reionization depends on the reionization history,
and thus is not universal at a given redshift.

Thus, it would be useful to attempt to quantify the effect of reionization
on the formation and evolution of the low-mass objects in a more complete
manner, relating the effective mass below which an object is a subject
to reionization feedback to the characteristic scales present at
each given moment of time.

\def\tableone{
\begin{deluxetable}{cccccc}
\tablecaption{Simulation Parameters\label{tabone}}
\tablehead{
\colhead{Run} & 
\colhead{$N$} & 
\colhead{Box size} & 
\colhead{Baryonic mass res.} & 
\colhead{Total mass res.} & 
\colhead{Spatial res.} }
\startdata
A & $128^3$ & $4h^{-1}{\rm\,Mpc}$ & $10^{5.7}\dim{M}_{\sun}$ & 
$10^{6.6}\dim{M}_{\sun}$ &$1.0h^{-1}{\rm\,kpc}$ \\
B & $128^3$ & $2h^{-1}{\rm\,Mpc}$ & $10^{4.8}\dim{M}_{\sun}$ & 
$10^{5.7}\dim{M}_{\sun}$ &$0.5h^{-1}{\rm\,kpc}$ \\
\enddata
\end{deluxetable}
}
\placefig{\tableone}
This paper attempts to accomplish precisely that based on the new simulations
of cosmological reionization. The 
simulations of a representative Cold Dark Matter cosmological 
model\footnote{With the following cosmological parameters: $\Omega_0=0.3$,
$\Omega_\Lambda=0.7$, 
$h=0.7$, $\Omega_b=0.04$, $n=1$, $\sigma_8=0.91$, where the amplitude and
the slope of the primordial spectrum are fixed by the COBE and cluster-scale
normalizations.}
were performed with the 
``Softened Lagrangian Hydrodynamics'' (SLH-P$^3$M) code (Gnedin 1995, 1996;
Gnedin \& Bertschinger 1996)
and fully described in Gnedin (2000).
Table \ref{tabone} lists two simulations used in this paper. Parameter $N$
gives the number of the dark matter particles; the quasi-Lagrangian baryonic 
mesh has the same size. Baryonic mass resolution is an average mass of a
baryonic cell, and the total mass resolution is the mass of a dark matter
particle plus the average mass of a baryonic cell. The spatial resolution
is measured as the gravitational softening length (the real resolution of
both the gravity solver and the gas dynamics solver is
a factor of two worse). Reionization by stars (i.e.\ with a soft UV 
background spectrum) is modeled with the Local Optical Depth approach,
which is able to approximately follow the three-dimensional radiative transfer
in the cosmological density distributions.

The two simulations from Table 1 allow me to investigate the sensitivity
of my results to the missing small and large scale power: run A has a larger
box whereas run B has a higher resolution. 
They also have different reionization histories, which allows me to test the 
generality of my results.

Both simulations
have sufficient mass resolution and the box size to resolve the relevant
characteristic mass scales during reionization and thus can be used
for the purpose of this paper. Since both simulations have
box sizes comparable to the nonlinear scale at the present time, they cannot
be continued until $z=0$. Rather, run A is stopped at $z=4$ and run B at
$z=6.5$.

The goal of this paper is to quantify the relationship between the
total mass of an object $M_{\rm t}$ and its gas mass $M_{\rm g}$.
The advantage of using the simulations listed in Table \ref{tabone} is that
they have enough resolution (both in mass and space) to actually map the
full two-dimensional distribution of objects in the $M_{\rm t}-M_{\rm g}$
plane. But before this can be done, I need to discuss what characteristic
mass scales are relevant for the evolution of the cosmic gas. This is
particularly important because, as the reader is reminded in the next section,
the Jeans mass, initially proposed as the characteristic scale, is essentially
irrelevant in the expanding universe.

\section{Reminder: the Linear Theory}

The effect of the reionization of the universe and the associated reheating
of the cosmic gas on the evolution of linear perturbations was comprehensively
discussed in Gnedin \& Hui (1998). As they showed, the relationship between
the linear overdensity of the dark matter $\delta_d(t,k)$ and the
linear overdensity of the cosmic gas $\delta_b(t,k)$ as a function of
time and the comoving wavenumber $k$, in the limit of small $k$ 
($k\rightarrow0$), 
can be written as
\begin{equation}
        {\delta_b(t,k)\over\delta_d(t,k)} = 1-{k^2\over k_F^2} +
	O(k^4),
        \label{defkf}
\end{equation}
where $k_F$ is in general a function of time. They called the physical scale
associated with the comoving wavenumber $k_F$ the {\it filtering scale\/},
since it is the characteristic scale over which the baryonic perturbations
are smoothed as compared to the dark matter.

The filtering scale is related to the Jeans scale $k_J$,
\begin{equation}
        k_J \equiv {a \over c_S}\sqrt{4\pi G\bar\rho}
        \label{defkj}
\end{equation}
(here $\bar\rho$ is the average total mass density of the universe, 
$c_S$ is the sound speed, which is uniquely defined in linear theory,
and $a$ is the cosmological scale factor), by
the following relation:
\begin{equation}
        {1\over k_F^2(t)} = {1\over D_+(t)} \int_0^t dt^\prime 
        a^2(t^\prime)
        {\ddot{D}_+(t^\prime)+2H(t^\prime)\dot{D}_+(t^\prime) 
        \over k_J^2(t^\prime)} 
        \int_{t^\prime}^t{dt^{\prime\prime}\over a^2(t^{\prime\prime})},
        \label{kfaskj}
\end{equation}
where $D_+(t)$ is the linear growing mode in a given cosmology.

For a flat universe at high redshift $z\ga2$, the scale factor $a$
is well approximated by the power-law in time, $a\propto t^{2/3}$,
and the growing mode $D_+$ is proportional to $a$. In this case equation 
(\ref{kfaskj}) can be substantially simplified:
\begin{equation}
        {1\over k_F^2(a)} = {3\over a} \int_0^a {da^\prime 
        \over k_J^2(a^\prime)} \left[1-\left(a^{\prime}\over a
	\right)^{1/2}\right].
        \label{kfaskjflat}
\end{equation}

Inspection of equation (\ref{kfaskj}) shows that the filtering scale {\it as a
function of time\/} is related to the Jeans scale {\it as a function of 
time\/}, but
at {\it a given moment in time\/} those two scales are unrelated and can be
very different. Thus, given the Jeans scale at a specific moment in time,
nothing can be said about the scale over which the baryonic perturbations
are smoothed. It is only when the whole time evolution of the Jeans
scale up to some moment in time is known that can the filtering scale at this
moment be
uniquely defined.\footnote{In general it is also true that the filtering 
scale is equal to the Jeans scale at some earlier moment in time.}

The physical explanation of this result is very simple: the gas temperature
(and thus the Jeans scale) can increase very quickly, but the gas density
distribution can only change on the dynamical time scale, which is about
the Hubble time for the linear perturbations. Thus, the effect of the
increased pressure on the gas density distribution will be delayed and can
only occur over the Hubble time.

While formally the filtering scale is only defined on large scales, as 
$k\rightarrow0$, the following formula
\begin{equation}
        {\delta_b(t,k)\over\delta_d(t,k)} \approx e^{-k^2/k_F^2}
        \label{approxkf}
\end{equation}
provides a remarkably accurate fit on all scales up to at least 
$k=k_F$, and is also very accurate when one needs to calculate the
integrals over the baryonic power spectrum $P_b(k)=\delta_b(t,k)^2$.
There is no obvious physical reason why equation (\ref{approxkf}) should be a
good fit to the full solution of the linear theory equations, but it was
extensively tested over a very large region of the parameter space
of possible cosmological models and
was always found to work well.

Since my task is to compare the baryonic versus the total mass
for small objects, it is convenient to switch from spatial to mass scales.
In linear theory, of course, there exists a one-to-one relationship between
the two. Thus, I can define the Jeans mass,
$$
	M_J \equiv {4\pi\over3}\bar\rho\left(2\pi a\over k_J\right)^3,
$$
and the filtering mass,
$$
	M_F \equiv {4\pi\over3}\bar\rho\left(2\pi a\over k_F\right)^3,
$$
as the mass enclosed in the sphere with the comoving 
radius equal to the corresponding spatial scale. The relationship between 
the two mass scales for $D_+\propto a\propto t^{2/3}$
can be easily obtained from equation (\ref{kfaskjflat}):
\begin{equation}
        M_F^{2/3} = {3\over a} \int_0^a da^\prime 
	M_J^{2/3}(a^\prime) \left[1-\left(a^{\prime}\over a
	\right)^{1/2}\right].
        \label{mfasmj}
\end{equation}

\section{Main Results}

\def\capTE{
The evolution of the mass- ({\it dotted line\/}) and volume-
({\it solid line\/}) weighted temperature from run A. Also shown is
the volume-averaged temperature for run B ({\it long-dashed line\/}).
The filled circles with error-bars label the virial temperature of 
objects that on average have the baryonic fraction of
50\% of the universal value. The triangles show
the Jeans temperature at the virial overdensity of 180
that corresponds to this virial temperature (eq.\ [\protect{\ref{tjdef}}]).
The right $y$ axis shows the circular velocity that corresponds to the
virial temperature marked by filled circles (the right $y$ axis has no meaning
for other curves on this plot).
}
\placefig{
\begin{figure}
\epsscale{0.65}
\insertfigure{\figdir/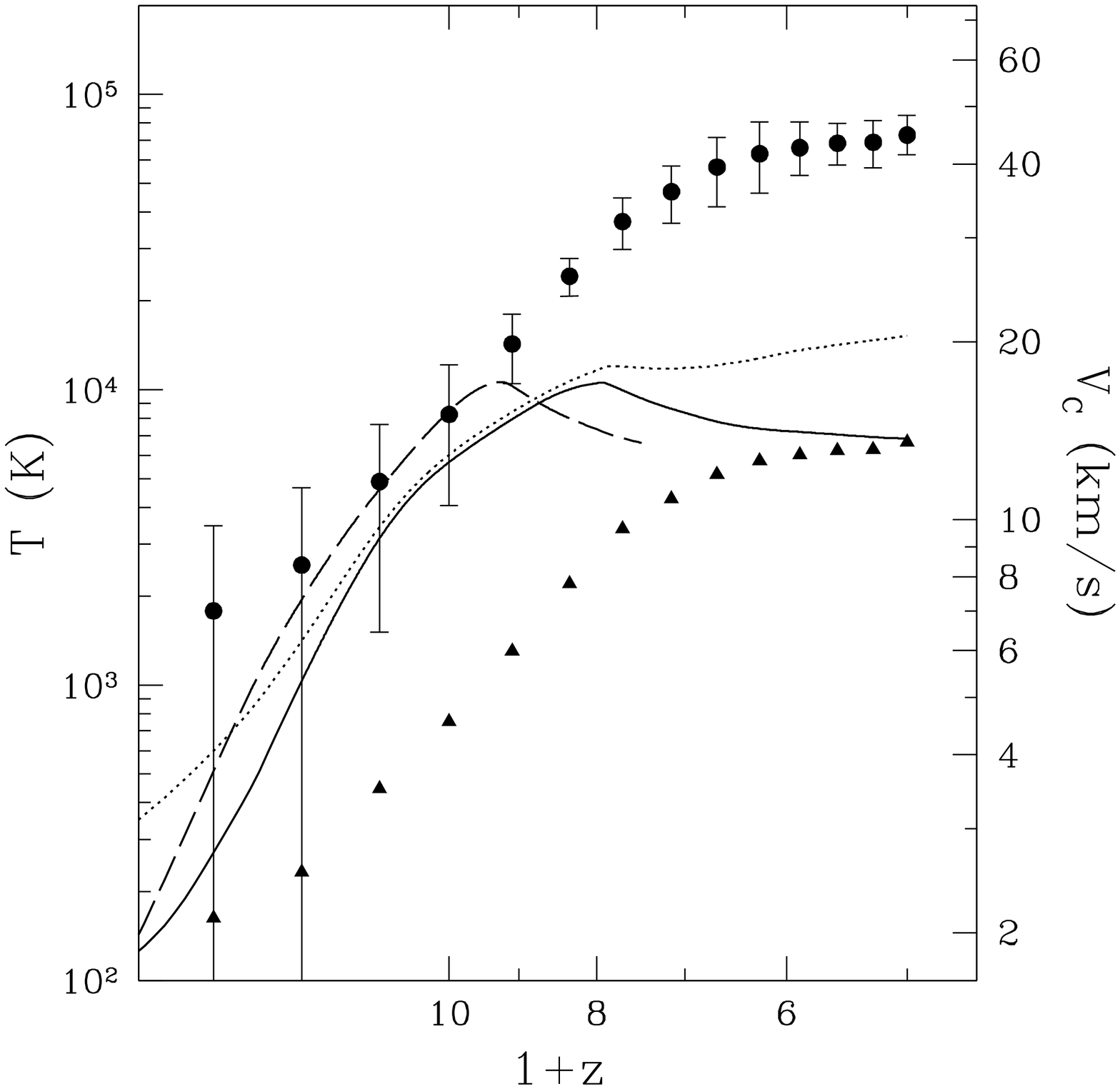}
\caption{\label{figTE}\capTE}
\end{figure}
}
Figure \ref{figTE} shows the evolution of the volume- and mass-weighted 
temperature in the simulations. Since the
gas density and temperature are not uniform in the simulation, the definition
of the linear sound speed (or linear temperature) is somewhat ambiguous.
I therefore adopt the volume-averaged temperature as a substitute for
the linear theory temperature, so that the linear theory sound speed,
which enters the definition of the linear theory Jeans mass, is defined as
$$
	c_S^2 = {5\over 3} {k_B\langle T\rangle_V\over \mu m_p},
$$
where $\mu=0.59$ is the mean molecular weight of the fully ionized gas.

However, the specific definition of the linear sound speed
is not very important. For example, if I used the mass-weighted
mean temperatures instead of the volume-weighted one, or instead, I
 calculate the mass- or 
volume-weighted Jeans mass directly from the simulation,
the difference in the computed
Jeans and filtering masses would be smaller than the statistical uncertainty 
due to a finite number of objects in my simulation (i.e.\ smaller than the
error-bars in Fig.\ \ref{figME}).

\def\capMM{
The gas versus the total mass for all objects from run A taken at
four different redshifts ({\it dark grey points\/}; redshift decreases
in the counter-clockwise direction). Also shown
with the light grey color the gas versus total mass at $z=15$.
The straight line marks the position of the universal baryon
fraction, $M_{\rm g}=(\Omega_b/\Omega_0) M_{\rm t}$, and three
curved lines show the fit to the function $\overline{M}_{\rm g}(M_{\rm t})$
together with 95\% confidence levels. The inserted axes show the
circular velocity corresponding to the total mass at each redshift.
}
\placefig{
\begin{figure}
\epsscale{0.70}
\insertfigure{\figdir/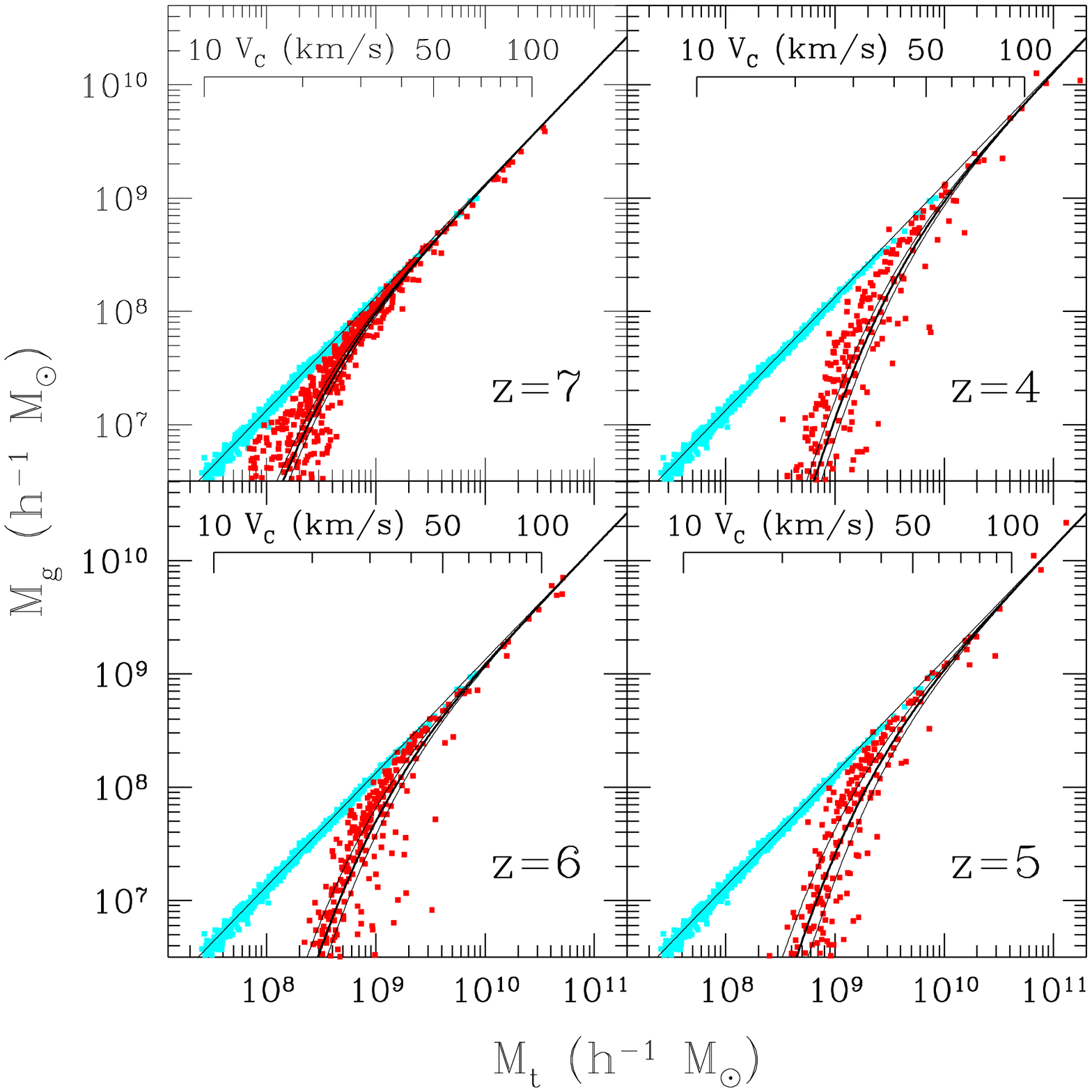}
\caption{\label{figMM}\capMM}
\end{figure}
}
The relationship between the gas and the
total mass of cosmological objects in the simulations
is now shown in Figure \ref{figMM}.\footnote{In all 
cases the stellar mass
makes only a small correction to the total gas mass and is 
ignored.} Bound objects are identified using the DENMAX algorithm of
Bertschinger \& Gelb (1991) with the gaussian density smoothing length
equal to one fifth of the mean interparticle separation, which 
corresponds to the characteristic overdensity of about 100.

Before reionization ($z=15$, light grey color) the gas
mass is directly proportional to the total mass of all objects, and the
coefficient of proportionality is the universal baryon fraction
$f_b\equiv\Omega_b/\Omega_0$. In the simulation shown (run A) 
reionization occurs
at $z\approx7$, and at that redshift the effects of reheating start to appear.
As time progresses, larger and larger mass objects are affected by the
increase in the gas temperature.

Without additional analysis, it is impossible to say whether the
reduction of the gas mass in low mass objects is due to the expulsion
of the already accreted gas, or due to the suppression of the accretion.
It is likely that both effects play a role - for example, essentially all
objects with masses below $2\times 10^8{\rm M}_{\sun}$ lost their gas
between $z=7$ and $z=6$, and since the average mass of a cosmological object
does not increase significantly during this short time interval, it is
clear that the gas was expelled from the low-mass objects. 

The first quantity of interest is the average baryonic mass of all objects
with given total mass, $\overline{M}_{\rm g}(M_{\rm t})$. This quantity
would be useful for semi-analytical modeling since it can be directly
plugged into the Press-Schechter approximation. Before reionization this
function has a very simple form, $\overline{M}_{\rm g}=f_b M_{\rm t}$, but
after reionization the small mass end of $\overline{M}_{\rm g}$ is suppressed.
In order to obtain a practically useful result, I approximate the mean 
baryonic mass with the following fitting formula,
\begin{equation}
	\overline{M}_{\rm g} = {f_b M_{\rm t}\over 
	\left[1+(2^{1/3}-1)M_C/M_{\rm t}\right]^3},
	\label{mcdef}
\end{equation}
which depends on a single parameter - the characteristic mass $M_C$, which is
the total mass of objects that on average retain 50\% of their gas mass.

In order to measure $M_C$ from the simulation, I first measure 
the average gas mass and its error-bars 
as a function of the total mass
from the simulation by 
averaging the gas mass of all objects within 0.1 dex around the
given value of the total mass. 
Then the value of $M_C$ and the corresponding error-bars are found by a
standard $\chi^2$ minimization.

The rationale for the particular choice of the fitting function is the 
following: it is clear from Fig.\ \ref{figMM} that at small masses the
mean baryonic mass goes as $M_{\rm t}^4$. I have therefore tried
fitting formulae of the following kind,
$$
	\overline{M}_{\rm g} = {f_b M_{\rm t}\over 
	\left[1+(2^{\alpha/3}-1)\left(M_C/M_{\rm t}\right)^\alpha
	\right]^{3/\alpha}},
$$
and $\alpha=1$ gives the best values for the $\chi^2$ test over the whole
time evolution.\footnote{At selected moments in time an equally good fit
can be obtained with different $\alpha$. For example, at $z=4$, $\alpha=2$
gives as good a fit as $\alpha=1$. By eye it appears that $\alpha=2$
might be better than $\alpha=1$, but this appearance is misleading;
Fig.\ \ref{figMM} is shown
on log-log scale, while $\overline{M}_{\rm g}$ is the average mass, and not
the exponent of the average logarithm of mass.}

\def\capME{
The evolution of various mass scales for two simulations: run A
({\it a\/}) and run B ({\it b\/}). The two thin lines show from the top
down the linear theory Jeans mass ($M_J$)
and the Jeans mass at the virial overdensity of 180 
($M_J/\sqrt{180}$). The bold line shows the filtering mass ($M_F$), and the 
symbols
with error-bars show the characteristic mass $M_C$
at which $\overline{M}_{\rm g} = 0.5 f_b M_{\rm t}$,
 as measured from the 
simulations.
}
\placefig{
\begin{figure}
\inserttwofigures{\figdir/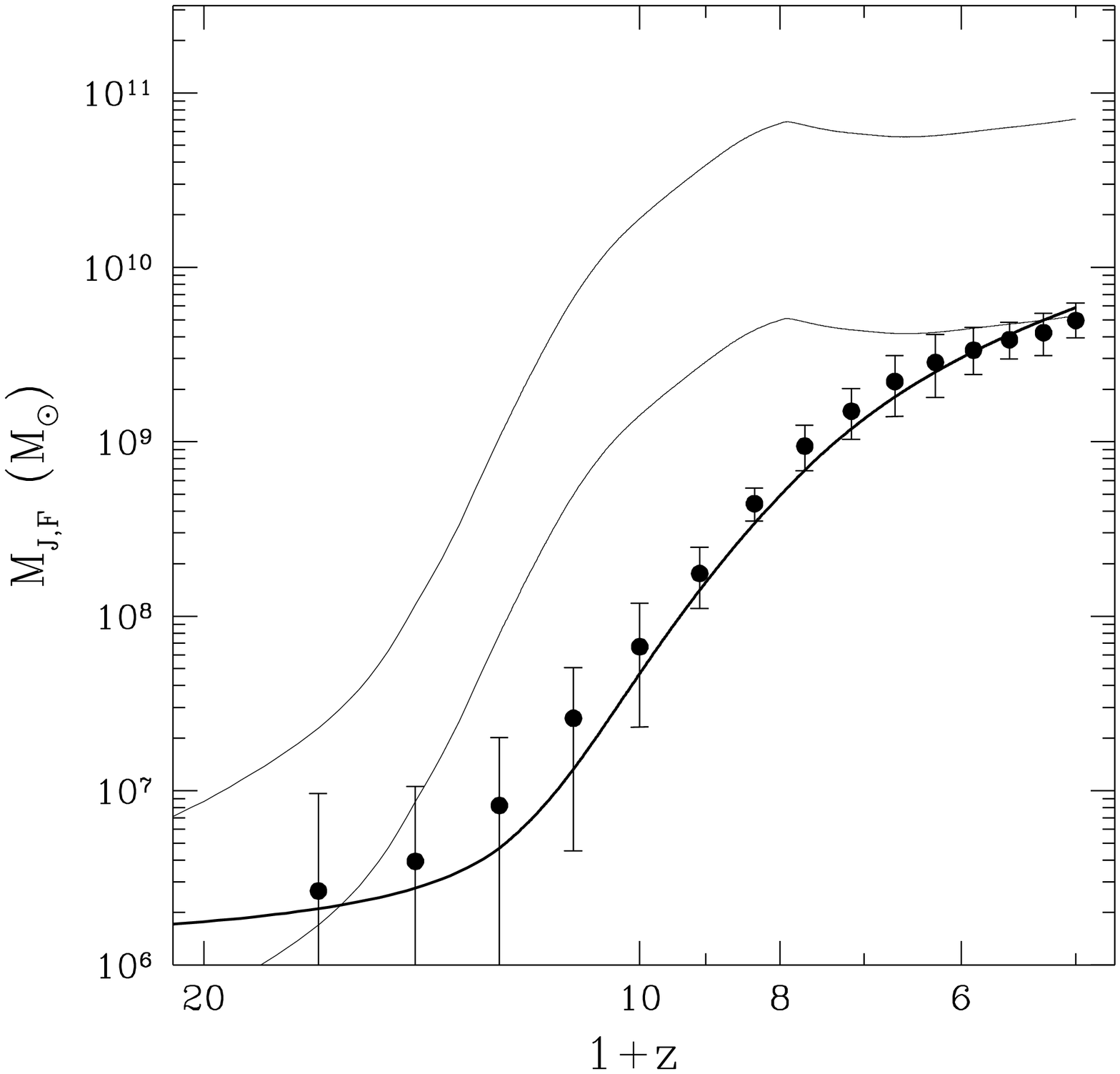}{\figdir/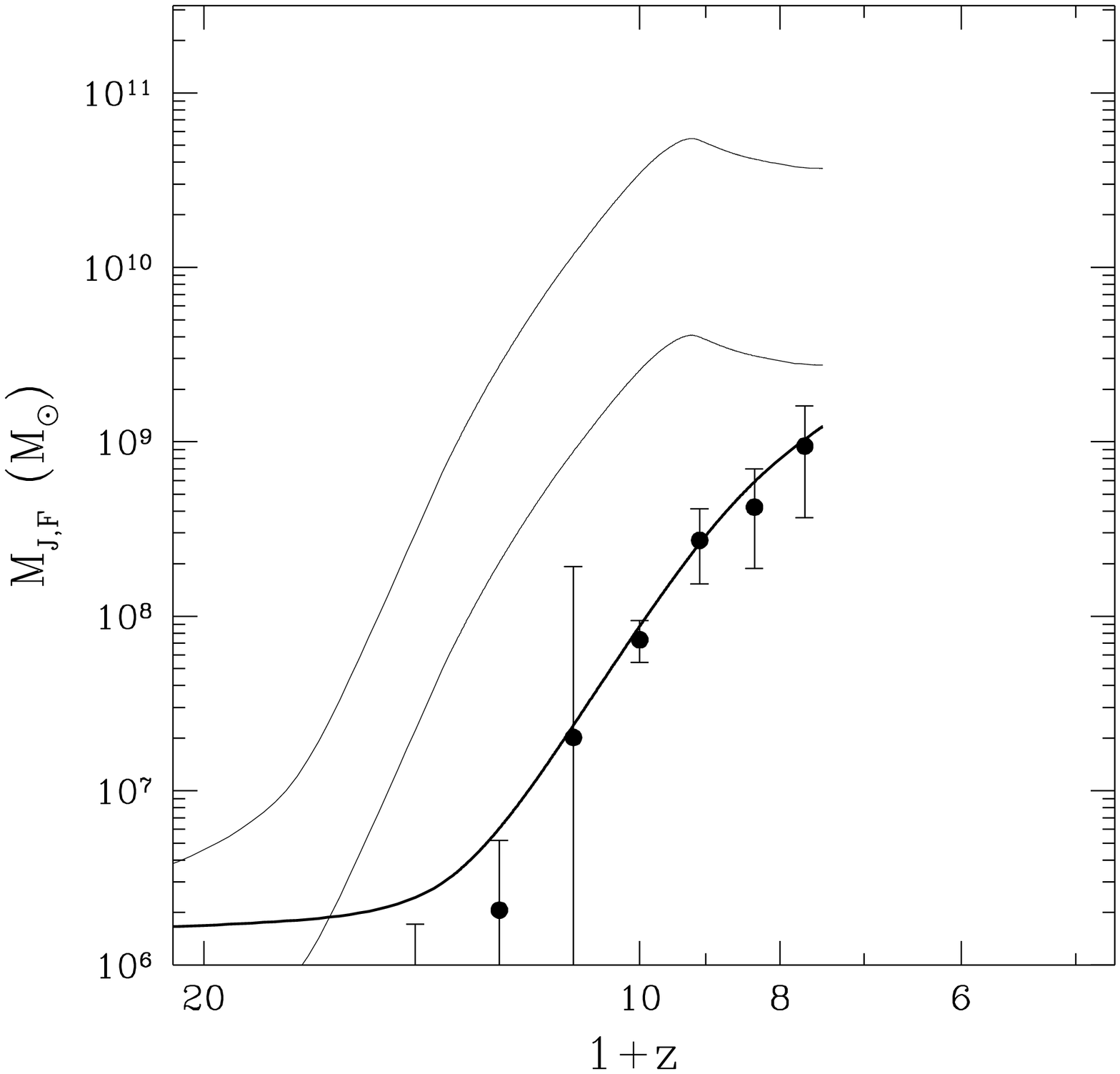}
\caption{\label{figME}\capME}
\end{figure}
}
Figure \ref{figME} now shows the evolution of the Jeans mass $M_J$, the
filtering mass $M_F$, and the characteristic mass $M_C$ for both simulations.
One can immediately see that the filtering mass provides a good fit for
the characteristic mass $M_C$,
whereas the linear theory Jeans mass is much larger.
One could, however, argue that the linear theory Jeans mass is the wrong scale
to use, and that the Jeans mass at the virial overdensity
$1+\delta=18\pi^2$ should be used instead. 
The lower thin line shows the virial overdensity Jeans mass, which provides a 
good
fit to the characteristic scale at the end of run A.
However, and this is the most important point of this paper,
{\it the overall shape is 
wrong\/}, which indicates that the agreement between the Jeans mass at the 
virial density and the characteristic mass scale at $z\sim4$ is a mere
coincidence. Rather, a linear theory filtering mass should be used
as an 
approximation to the characteristic mass $M_C$,
\begin{equation}
	\overline{M}_{\rm g}(M_{\rm t},t) \approx {f_b M_{\rm t}\over 
	\left[1+(2^{1/3}-1)M_F(t)/M_{\rm t}\right]^3}.
	\label{fitfor}
\end{equation}

This result can also be cast in the temperature units. The circles with
the error-bars in Fig.\ \ref{figTE} show the virial temperature,
$$
	T_{\rm vir} \equiv {\mu m_p\over 2 k_B} v_c^2 =
	{\mu m_p\over k_B} G M_{\rm t}^{2/3} (3\pi^3\bar\rho)^{1/3}
$$
(Thoul \& Weinberg 1996) that corresponds to the characteristic mass $M_C$
for run A (the right $y$ axis shows the respective circular velocity).
The same mass corresponds to the 
Jeans mass at the virial overdensity for 
gas with the temperature of
\begin{equation}
	T_J \equiv {\mu m_p\over k_B}{3^{7/3}\over 10\pi}
	G M^{2/3} \bar\rho^{1/3} = 
	{9\over 10\pi^2} T_{\rm vir}.
	\label{tjdef}
\end{equation}
The value of $T_J$ is shown in 
Fig.\ \ref{figTE} with the filled triangles.
Thus, if one wants to use the
Jeans mass at the virial overdensity to quantify the effect of 
reionization on structure
formation, one must use the effective temperature $T_{\rm eff}$ of
reheated gas so that $M_{J,\rm eff} = M_F$,
which in terms of temperature translates into
$T_{\rm eff} \sim 8\times10^3 \dim{K}$
at $z\sim4$ and $T_{\rm eff} \sim 2\times10^3 \dim{K}$ during reionization,
rather that $T_{\rm eff}=(1-2)\times10^4\dim{K}$ at all times.

Thus, the effect of the expansion of the
universe (which causes the delay of the growth of the
filtering mass $M_F$ with respect to the linear Jeans mass $M_J$)
on the gas fraction suppression in low mass objects
has the same appearance as the non-expanding universe with the
reheating temperature of only a few thousand degrees.

For most practical applications the knowledge of 
$\overline{M}_{\rm g}(M_{\rm t},t)$
may be sufficient, but the resolution of my simulations is also sufficient
to approximately characterize the whole distribution function
$P(M_{\rm g},M_{\rm t})$. Using the product rule of probability theory,
this distribution function can be written as the product of the
probability to find an object with the total mass $M_{\rm t}$
and the probability distribution of the gas masses for all
objects with a given total mass,
$$
	P(M_{\rm g},M_{\rm t}) = P(M_{\rm t})P(M_{\rm g}|M_{\rm t}).
$$
The former probability depends on the cosmological model, whereas the latter
may be conjectured to be independent of the specific model, and
depend only on the characteristic mass scales present at this moment.
The physical reason for this conjecture is simple: the number density
of cosmological objects of a given mass depends on the cosmological model,
but as long as this number density is not exceedingly high, different objects
as virialized entities are independent of each other and the rest of
the universe, and thus a probability for an object to lose a given fraction
of its gas mass should not depend strongly on the abundance of other objects
somewhere else in the universe.

The number of objects in the simulations is not high enough
to accurately measure the probability distribution function
$P(M_{\rm g}|M_{\rm t})$, but it appears that for any $M_{\rm t}$ and 
at any given
time this distribution is compatible with a lognormal distribution.
Assuming that it is indeed lognormal, I can write it down as follows:
\begin{equation}
	P(M_{\rm g}|M_{\rm t}) = {1\over \sigma \sqrt{2\pi}}
	\exp\left[-{1\over2\sigma^2}\left(\ln M_{\rm g} - 
	\ln\overline{M}_{\rm g}+\sigma^2/2\right)^2\right],
	\label{ln}
\end{equation}
where $\sigma$ is the rms dispersion of the logarithm of the gas mass,
and is a function of the total mass, as is $\overline{M}_{\rm g}$
(eq.\ [\ref{fitfor}]).

\def\capMS{
The same as Fig.\ \protect{\ref{figME}}, except that the symbols now
show the characteristic mass $M_C^*$ from the fitting formula for the
rms dispersion $\sigma$ of the lognormal distribution 
$P(M_{\rm g}|M_{\rm t})$
at a given $M_{\rm t}$.
}
\placefig{
\begin{figure}
\inserttwofigures{\figdir/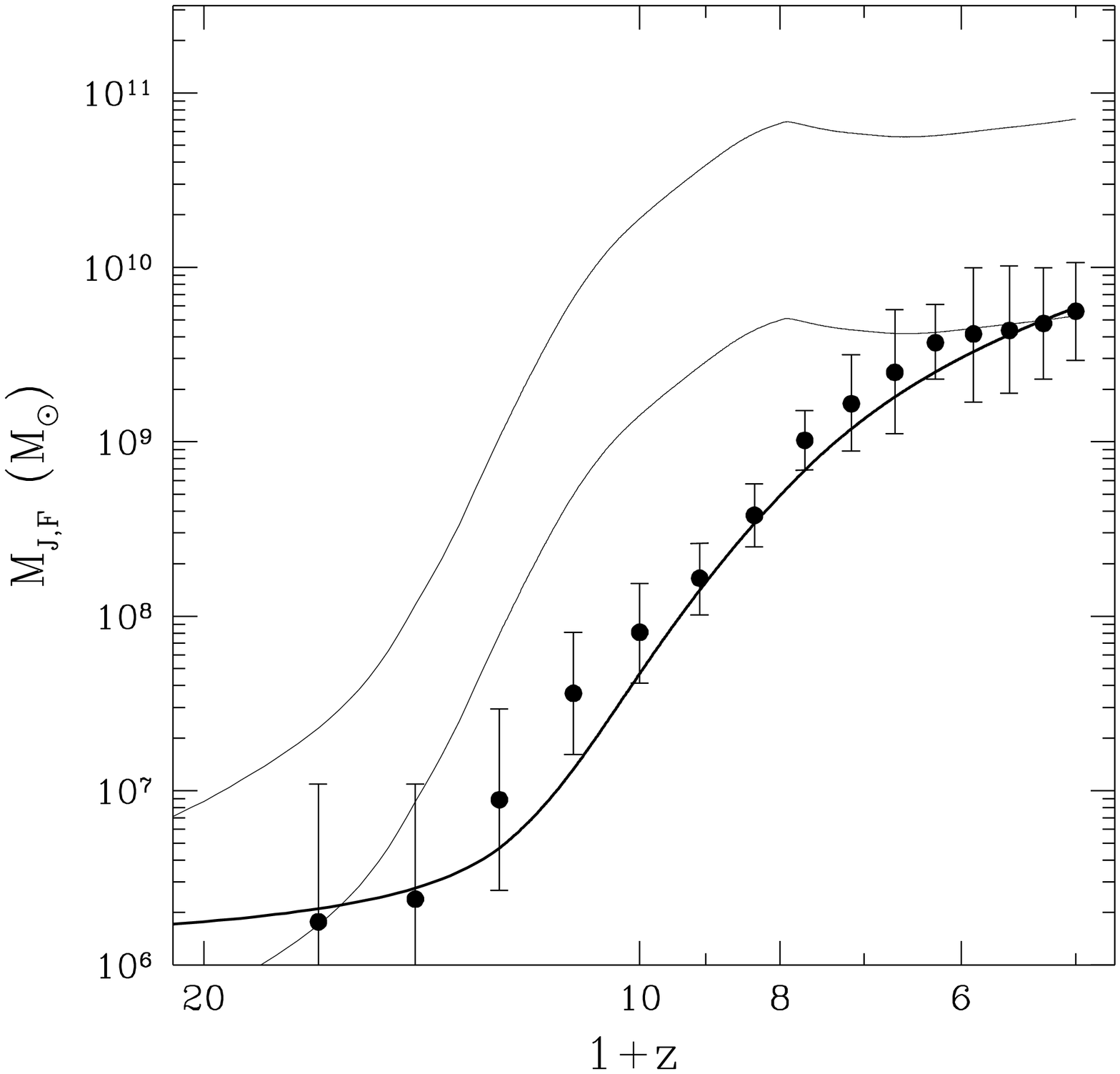}{\figdir/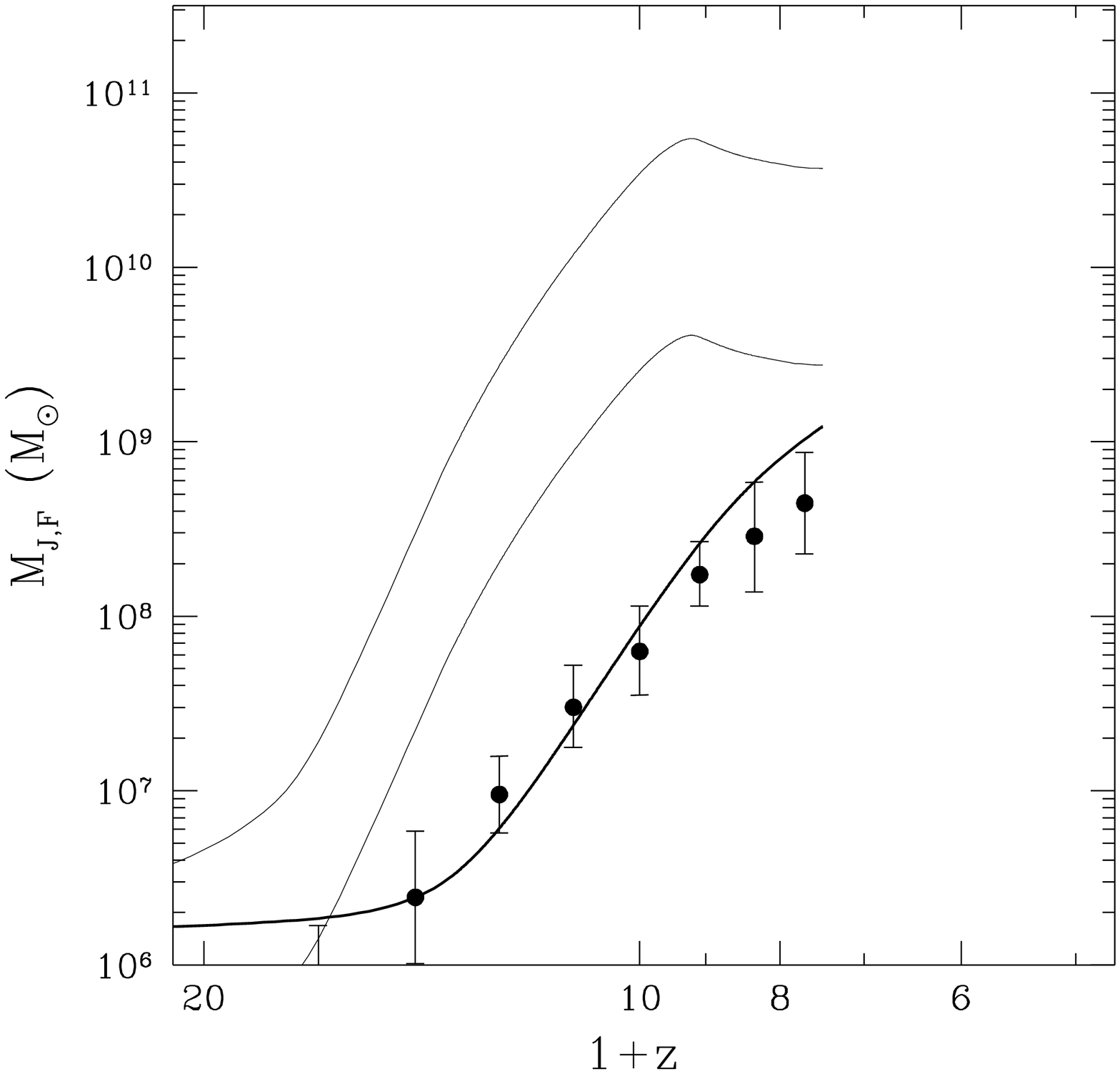}
\caption{\label{figMS}\capMS}
\end{figure}
}
In order to come up with a closed form fitting formula, I approximate $\sigma$
by the following power-law:
$$
	\sigma(M_{\rm t}) = {M_C^*\over 3 M_{\rm t}},
$$
where $M_C^*$ is another characteristic mass, which is plotted in Figure
\ref{figMS} together with the Jeans mass and the filtering mass. Again,
as one can see, the characteristic mass $M_C^*$ is approximately equal
to the filtering mass, so that the rms dispersion of the logarithm of
gas mass at a given total mass is given by the following fit:
\begin{equation}
	\sigma(M_{\rm t},t) \approx {M_F(t)\over 3 M_{\rm t}}.
	\label{sigfit}
\end{equation}
The last equation is not very well constrained by my simulations. For example,
the coefficient in the denominator is determined with only 30\% accuracy,
so replacing it with 2 or 4 also gives an acceptable fit to the data.

Equations (\ref{fitfor}-\ref{sigfit}) give in a closed form the probability
distribution function
$P(M_{\rm g}|M_{\rm t},t)$ for any cosmological model and 
reionization history (which are specified through the filtering mass
$M_F(t)$ as a function of time).

\section{Low Redshift Evolution}

\def\capMF{
({\it a\/})
Extrapolation of Fig.\ \protect{\ref{figME}a} to $z=0$. Lower solid curves
show the extrapolation assuming the temperature
evolution $T\propto a^{-0.88}$, whereas the upper curves
also include a second reheating at $z=3$ as indicated by the
observations. ({\it b\/}) The same plot with masses converted into circular 
velocities. The triangle shows the results of Thoul \& Weinberg (1996) and
Quinn et al.\ (1996) for the model without reheating (lower solid curves).}
\placefig{
\begin{figure}
\inserttwofigures{\figdir/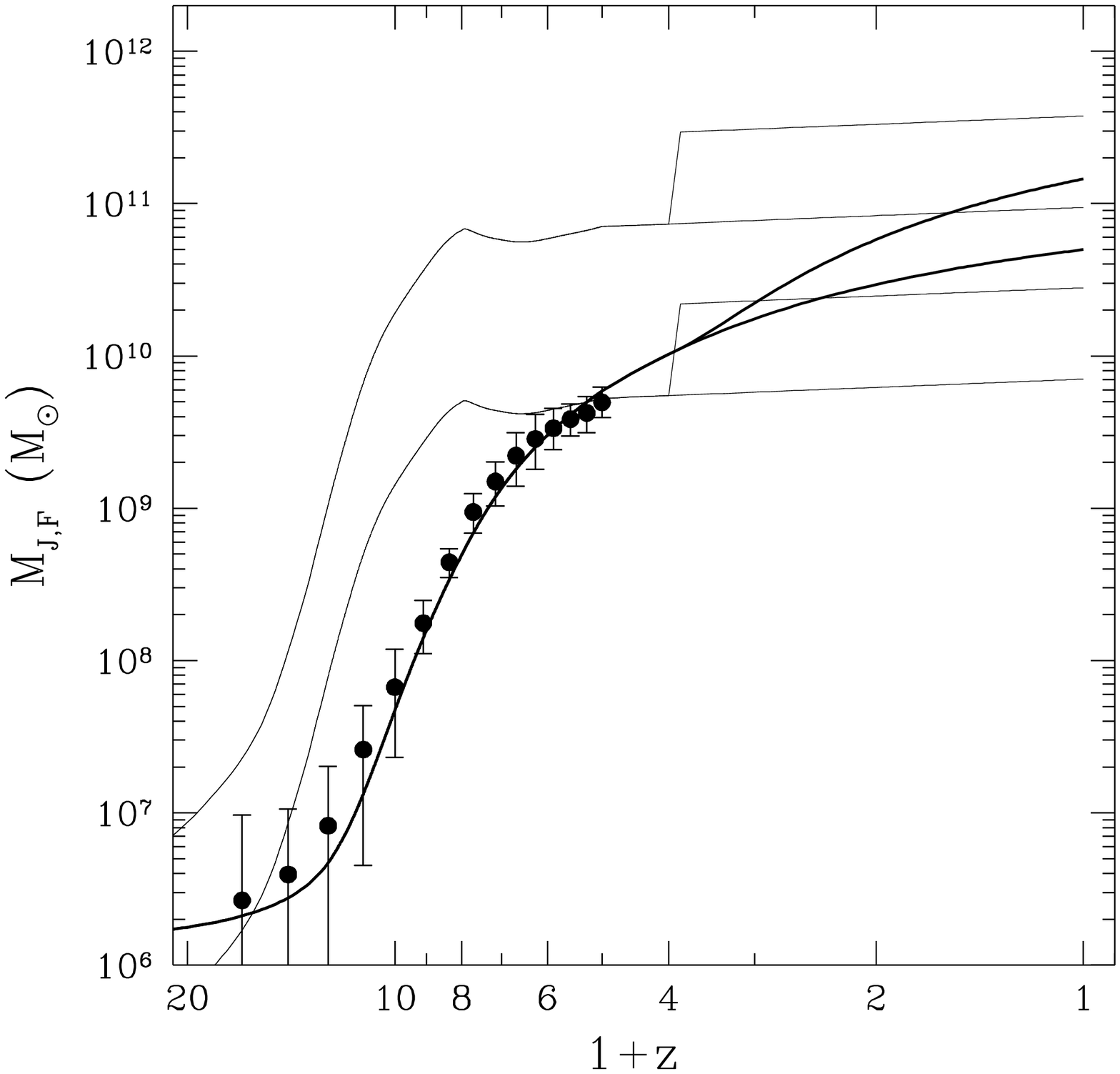}{\figdir/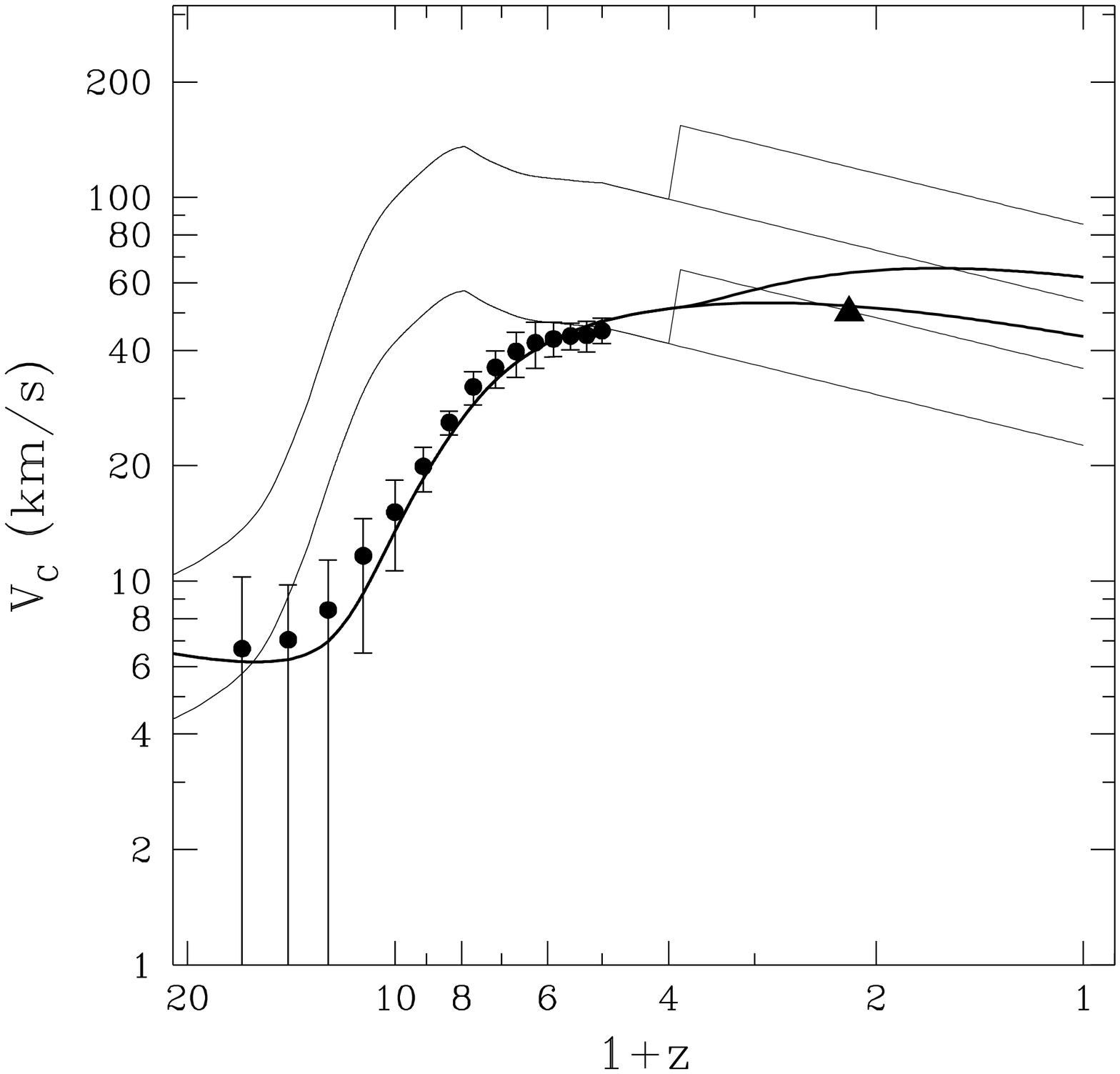}
\caption{\label{figMF}\capMF}
\end{figure}
}
Given the thermal history of the universe, the evolution of the linear theory
Jeans mass and thus the filtering mass can be calculated up to the present
time. In a universe with a single (hydrogen) reionization epoch
the temperature
at late times is theoretically predicted to evolve as $\propto a^{-0.88}$
(Miralda-Escud\'{e} \& Rees 1994; Hui \& Gnedin 1997). However, 
recent analyses of Lyman-alpha forest data indicate that the universe
underwent a second reheating at $z\approx 3$ (Ricotti, Gnedin, \& Shull
2000; Schaye et al.\ 2000), possibly as the result of helium reionization
by quasars.
In order to account for it, I also consider
a thermal history for which the gas temperature is increased sharply
at $z=3$ by a factor of 2.5. Figure \ref{figMF} shows the predicted
evolution of the filtering mass, the linear theory Jeans mass, and the
Jeans mass at the virial overdensity for the two cases (the upper curves
correspond to the secondary reheating model) in an $\Omega_0=1$ universe
(for a flat universe with the cosmological constant, the result is almost
indistinguishable). This paper therefore makes a 
prediction that objects as massive as $10^{11}{\rm M}_{\sun}$ have on average
a baryon fraction of only 50\% of the universal value. (The baryonic
content of these objects is likely to be dominated by stars at the present
time.) Note, however, that in terms of circular velocities, 
the characteristic scale is essentially independent of redshift, and is
about $50\dim{km/s}$, 
in agreement with previous investigations (Thoul \& Weinberg 1996;
(Quinn, Katz, \& Efstathiou 1996; Weinberg, Hernquist, \& Katz 1997). 
The Jeans mass
at the virial overdensity is, however, a decreasing function of time,
and corresponds to a circular velocity of only about $30\dim{km/s}$
at $z=2$ in the model without the secondary reheating, contrary to the 
numerical results. This gives a
further illustration of the fact that the Jeans mass (at any
overdensity) is not the proper scale that controls the gas fraction
in low mass objects.

\section{Conclusions}

Based on cosmological simulation of reionization, I showed that the 
linear Jeans
mass does not control the mass scales over which reionization suppresses
the gas fraction in low mass cosmological objects, and may overestimate
the characteristic mass scale by an order of magnitude.  
Since the Jeans mass is
a function of the density, at the virial overdensity of 180 the
Jeans mass is able to match the characteristic scale at later times,
but the general shape of the Jeans mass vs redshift and the characteristic
mass vs redshift do not match. 
Instead, the
filtering mass, which directly corresponds to the length scale over which
the baryonic perturbations are smoothed in linear theory, provides
a good fit to the characteristic mass scale. This conclusions supports
a very simple picture (proposed by Shapiro et al.\ 1994)
in which the effect of reionization on structure
formation in the universe is controlled by a single mass scale
both in the linear and the nonlinear regime. However, this work 
demonstrates that it is not the Jeans mass (as was previously thought)
but the filtering mass that is the approriate mass scale.

The distribution of the
gas fractions of all cosmological objects with the same total mass at any
given moment during the evolution of the universe
is approximately lognormal, is fully specified by the filtering mass
at this moment, and is given by equations (\ref{fitfor}-\ref{sigfit}).

These equations, supplemented with the total-mass function of cosmological
objects, fully describe the full probability distribution to find an object
with given values of its gas and total mass. This probability distribution
can be conveniently used in semi-analytical modeling of the evolution
of low mass objects in the universe.

Admittedly, the simulations used in this paper consider only one cosmological
model, but all cosmological models (including those that are
curvature- or vacuum-dominated today) have an expansion law that approaches
$a\propto t^{2/3}$ at sufficiently
high redshift, and follow similar behavior. Thus, while the 
applicability of equations (\ref{fitfor}-\ref{sigfit}) for a wide range of
cosmological models is not rigorously proven, the physical reasoning
suggests that this is likely to be the case.

My simulations also do not cover all the possible reionization histories, but
at least runs A and B have different reionization histories and different
resolutions, which gives some credibility to the conjecture that equations
(\ref{fitfor}-\ref{sigfit}) work for different reionization histories and are
also free from numerical artifacts.

\acknowledgements

I am grateful to the referee David Weinberg for fruitful comments that
substantially improved the original manuscript.
This work was partially supported by National Computational Science
Alliance under grant AST-960015N and utilized the SGI/CRAY Origin 2000 array
at the National Center for Supercomputing Applications (NCSA).

\placefig{\end{document}}

\clearpage

\newcounter{figurecap}
\setcounter{figurecap}{0}

\begin{center}
\bf Figure Captions
\end{center}

\refstepcounter{figurecap}
Fig.\ \thefigurecap---\label{figTE}\capTE

\refstepcounter{figurecap}
Fig.\ \thefigurecap---\label{figMM}\capMM

\refstepcounter{figurecap}
Fig.\ \thefigurecap---\label{figME}\capME

\refstepcounter{figurecap}
Fig.\ \thefigurecap---\label{figMS}\capMS

\refstepcounter{figurecap}
Fig.\ \thefigurecap---\label{figMF}\capMF

\clearpage

\tableone

\end{document}